\documentclass[journal]{IEEEtran}

  \ifCLASSINFOpdf

  \else

  \fi

\usepackage{multirow}

  \usepackage{amsfonts}
  \usepackage{graphicx}
  \graphicspath{ {images/} }
  \usepackage{caption}
  \usepackage{subcaption}
  \usepackage{cite}
  \usepackage{graphicx}
  \usepackage{psfrag}
  \usepackage{url}
  \usepackage{stfloats}
  \usepackage{amsmath}
  \usepackage{array}
  \usepackage{amssymb}
  \usepackage{setspace}
  \usepackage{algorithmicx}
  \usepackage[ruled]{algorithm}
  \usepackage{algpseudocode}
  \usepackage{algpascal}
  \usepackage{algc}
  \usepackage{color}

\usepackage{url}
\usepackage{lipsum}
\usepackage{blkarray}
\usepackage{xcolor}
\usepackage{soul}
\usepackage{booktabs}
\usepackage{multirow}
\usepackage{microtype}
\usepackage{epstopdf}

\begin{document}

  \title{\LARGE Rate-Loss Mitigation for a Millimeter-Wave Beamspace MIMO Lens Antenna Array System Using a Hybrid Beam-Selection Scheme}

  \author{Md. Abdul Latif Sarker, Md. Fazlul Kader, and Dong Seog Han\vspace{-4ex}
  \thanks{This work was supported by the Brain Korea 21 (BK21) Plus project funded by the Korean Ministry of Education (21A20131600011), a Korea Institute for Advancement of Technology (KAIT) grant funded by the Korean Government (No.P0000535) and in part by Samsung Electronics Co., Ltd under 802.11ad standard development project for 60-GHz millimeter-wave frequency band. (\textit{corresponding author: Dong Seog Han.})}
  \thanks {M. A. L. Sarker and D. S. Han are with the School of Electronics Engineering, Kyungpook National University, Daegu 41566, South Korea. (e-mail: latifsarker@jbnu.ac.kr; dshan@knu.ac.kr).}
  \thanks {M. F. Kader is with the Department of Electrical and Electronic Engineering, University of Chittagong, Chittagong 4331, Bangladesh (email: f.kader@cu.ac.bd).}
   }

  \markboth{IEEE Systems Journal,~Vol.~xx, No.~xx, xx~2019}%
  {}


  \maketitle

  \begin{abstract}
    The rate loss is an important issue for multi-user beamspace multiple-input multiple-output (MIMO) millimeter-wave systems. In such systems, multi-cluster channel propagation leads as a function of the quantization error and, as a result, each active user suffers a significant rate loss. Particularly, a conventional single beam selection (SBS) scheme is used in an equivalent beamspace MIMO channel to reduce the required number of radio frequency (RF) chains, but this scheme does not properly mitigate the quantization error due to the limited number of the channel propagation clusters per-user. In addition, optimally selected beams or RF chains are operated at the same number of the precoded data streams. Consequently, each active user inherently loses a huge number of data streams utilizing the conventional SBS scheme. Therefore, we propose a hybrid beam selection (HBS) scheme in an equivalent beamspace MIMO channel to implement a multiple beam or RF chain group selection opportunity and choose a reliable channel propagation cluster per-active-user, which directly affects to mitigate the quantization error. The numerical result is verified by computer simulations in terms of a millimeter-wave MIMO downlink channel environment.
  \end{abstract}
  \IEEEpeerreviewmaketitle
  \section{Introduction}
 \IEEEPARstart{M}{illimeter-wave} (mmWave) beamspace multiple-input multiple-output (MIMO) communications represent promising technology for 5th-generation wireless communication and have received great attention owing to their large bandwidth availability \cite{ 1}. Such systems use the 30-300 GHz frequency band, while microwave wireless communications operate at carrier frequencies below 6 GHz. A traditional MIMO technology is implemented digitally at the baseband
 ~\cite{2}, which may not be useful in mmWave beamspace MIMO systems because of its high hardware cost.
 Accordingly, analog beamforming implemented \cite{3} 
to decrease costs via phase shifters in the RF frontend. Additionally, an antenna subset selection algorithm was proposed in \cite{4} to replace phase shifters with switches. Another study proposed a hybrid analog-digital precoding architecture in \cite{5, 6}.
In \cite{5}, the authors exploited the spatial structure of mmWave channels to formulate the precoding/combining problem. A unified heuristic design is implemented in [6]. 
Thus, the application of mmWave MIMO systems is more challenging due to unpredictable propagation environments, system efficiency and power consumption.\par

 Recently, several researchers investigated prototype lens antenna arrays (LAA) in MIMO systems~\cite{7, 8, 9, 10, 11} to generate a virtual phase shifter and design a confined beamspace channel. Particularly, An LAA prototype treats the electromagnetic lens as a `sinc' function and an antenna array to enable a MIMO system in \cite{7}. 
 A path division multiple-access method with a full-dimensional LAA 
was proposed in \cite{8} where the method followed two `sinc' functions reflecting both the elevation and the azimuth angle resolutions.
 In~\cite{9}, authors investigated an overview of signal processing techniques with LAA scheme for mmWave communications. In the most recent studies \cite{ 10, 11}, an existing single beam selection (SBS) scheme is investigated in an equivalent beamspace MIMO channel. An SBS scheme with a support detection algorithm is applied to reduce the pilot overhead in \cite{10}. Authors considered an SBS scheme with a finite-rate feedback-based hybrid zero-forcing precoding to minimize the MIMO dimension and analyze the quantization error (QE) in \cite{11}. 
 However, most of the relevant existing literature has followed the SBS scheme to reduce the number of the RF chains, but this setup does not make a multiple beam or RF chain group selection opportunity for each active user.  Therefore, we propose a hybrid beam selection (HBS) scheme to implement a multiple beam or RF chain group selection opportunity and choose a suitable channel cluster per-active-user, which resort to significantly enhance the data rate per-user. To the best of our knowledge, this issue has not yet been investigated in the literature.\par

In this letter
, we propose an HBS scheme in an equivalent beamspace MIMO downlink channel for mmWave LAA system. It should be noted that developing a better beam selection scheme to choose a reliable beam or RF chain group and sort out a suitable channel cluster for every supported user, is a main factor of this letter. We verify the superiority of the proposed HBS scheme over the SBS scheme, in terms of each user data rate performance through computer simulations.

 \section{Proposed System Model}
 We consider a single-cell downlink massive mmWave MIMO system where the base station (BS) is equipped with $M$ transmitting 
 antennas. In this system, $K$ single-antenna users are simultaneously served by the BS equipped with $N_{RF}$ RF chains, where $M\gg K$ and 
 $K\leq N_{RF}\leq M$ in \cite{8, 9, 10, 11}. The system model of the beamspace MIMO is depicted in Fig. 1 where the existing system model is described in Fig. 1(a) and the proposed system model is presented in Fig. 1(b). From Fig. 1(b), we assume $N_{RF_{g}}=N_{RF_{g_1}}+N_{RF_{g_2}}$ and $\mathbf{S}_{h}=[\xi\mathbf{S}_{1} $  $(1-\xi)\mathbf{S}_{2}]$ where $N_{RF_{g_1}}$ is an RF chain group $1$, $N_{RF_{g_2}}$ is an RF chain group $2$, $\mathbf{S}_{h}$ is a $M\times N_{RF_{g}}$ hybrid beam selection (HBS) matrix, selector $\mathbf{S}_{1}$ is an $M\times N_{RF_{g_1}}$ matrix, selector $\mathbf{S}_{2}$ is an $M\times N_{RF_{g_2}}$ matrix, selector $\mathbf{S}_{1}$ and $\mathbf{S}_{2}$  contains one non-zero element `1' in every column with a different position and $0\leq\xi\leq 1$ denotes a constant and indicates the amount of directional transmission. It is noted that the designed HBS scheme does not manipulate the selector dimension, but it implements a multiple beam or RF chain group selection scope to find out a suitable channel propagation cluster for each active user. Next, we adopt a massive mmWave MIMO downlink channel vector, $\mathbf{h}_{k} \in {\mathbb{C}}^{M\times 1}$, between the BS antennas and the $k$-th user. The received signal, $r_{k}$, can therefore be modeled as
\begin{figure*}[htb]
\centering
\begin{subfigure}{.5\textwidth}
  \centering
\includegraphics[width=2.5in,height=2.5in, keepaspectratio]{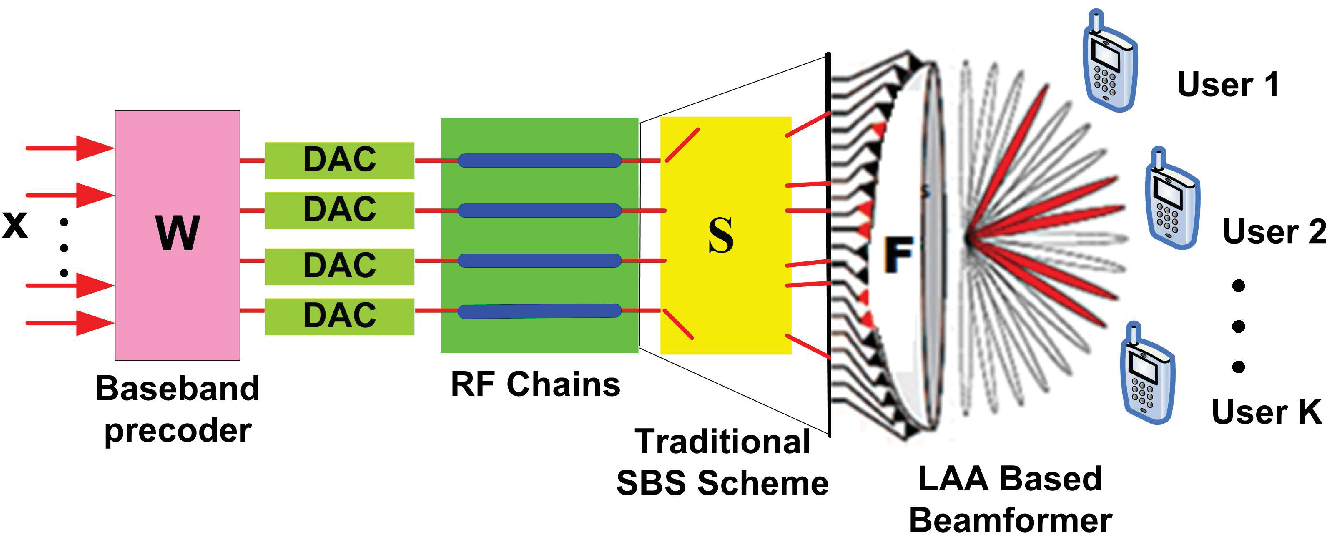}
  \caption{beamspace MIMO with traditional SBS scheme}
  \label{Fig1a}
\end{subfigure}%
\begin{subfigure}{.5\textwidth}
  \centering
\includegraphics[width=2.5in,height=2.5in, keepaspectratio]{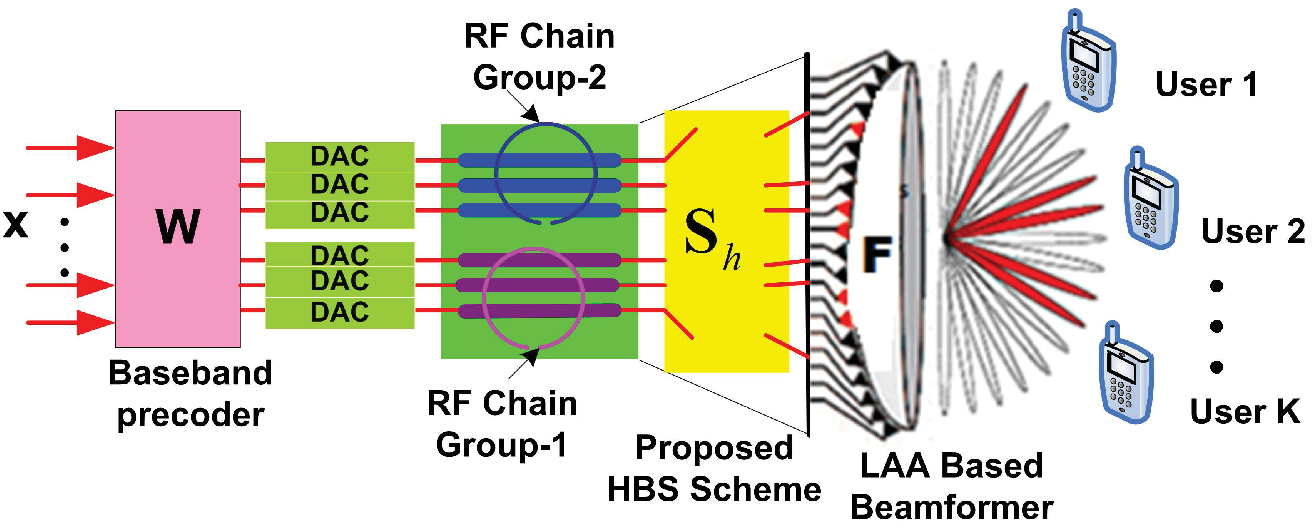}
  \caption{beamspace  MIMO with proposed HBS scheme}
  \label{Fig1b}
\end{subfigure}
\caption{System Model (a) beamspace MIMO with traditional SBS Scheme (b) beamspace  MIMO with proposed HBS scheme.}
\label{Fig1}
\end{figure*}
\raggedbottom

 \begin{equation}
 r_{k}=\sqrt\frac{\rho}{K}\mathbf{h}_{k}^{H}\mathbf{F}^{H}\mathbf{S}_{h}\mathbf{Wx}+n_{k},
 \end{equation}
 where $\mathbf{h}_{k}=\sum_{i}^{L_{k}}\alpha_{k,i}\textrm{\textbf{a}}(\phi_{k,i})$ \cite{9}, $L_{k}$ is the number of dominant channel paths of the $k$-th user (which is typically small), $L_{k}\ll N_{RF_{g}}$ and $L_{k}$ are no larger than 3 because of the multi-path sparsity of mmWave channels \cite{8, 9, 11, 12}, $\alpha_{k,i}$ denotes the complex-valued $i$-th path gain at the $k$-th user, $\textrm{\textbf{a}}(\phi_{k,i})$ is the $M\times 1$ steering vector of the $i$-th path at the $k$-th user, $\mathbf{F}=[\textrm{\textbf{a}}(0), \textrm{\textbf{a}}(\delta),...,\textrm{\textbf{a}}(\delta(M-1))]^{H}$ is the $M\times M$ spatial-domain discrete Fourier transform matrix with $\delta=\frac{1}{M}$~\cite{9, 10, 11}, the lens 
 is characterized by $\mathbf{F}$, $\mathbf{W}=[\textbf{w}_{1}, \textbf{w}_{2},...,\textbf{w}_{K}]$ is the $N_{RF_{g}}\times K$ zero-forcing (ZF) precoding matrix, $\rho$ is the transmit power, $\mathbf{x}$ is the $K\times 1$ data vector with normalized power, the $\mathbb{E}[\mathbf{x}\mathbf{x}^{H}]=\textbf{I}_{K}$, the $(\cdot)^{H}$ operator represents the Hermitian, and $n_{k}$ is the complex Gaussian noise with zero mean and unit variance at the $k$-th user. 
  The $M$-element uniform linear array (ULA) with $\phi_{k,i}=\frac{d}{\lambda}\textrm{sin}(\theta_{k,i})$ is treated as in \cite{10}, where the wavelength $\lambda=\frac{c}{f_{c}}$, $c$ is the light speed, $f_{c}$ is the carrier frequency, $d=\frac{c}{2f_{c}}$ is the antenna spacing, and $\theta_{k,i}$ represents the angle of departure (AoD) of the $i$-th path at $k$-th user. Thus, the steering vector is given by \cite{4} and \cite{8} as follows:
 \begin{equation}
 \textrm{\textbf{a}}(\phi_{k,i})=\frac{1}{\sqrt{M}}[1, e^{-j2\pi\phi_{k,i}},...,e^{-j2\pi\phi_{k,i}(M-1)}]^{T}.
 \end{equation}
 \section{Proposed Equivalent Beamspace MIMO Channel}
We consider the $M\times K$ beamspace channel matrix, $\mathbf{H}_{b}$, as
\begin{equation}
 \mathbf{H}_{b}=[\mathbf{F}\mathbf{h}_{1},\mathbf{F}\mathbf{h}_{2},...,\mathbf{F}\mathbf{h}_{K}]=[\mathbf{h}_{b,1},\mathbf{h}_{b,2},...,\mathbf{h}_{b,K}],
 \end{equation}
where $\mathbf{H}=[\mathbf{h}_{1},\mathbf{h}_{2},...,\mathbf{h}_{K}]\in {\mathbb{C}}^{M\times K}$,  $\mathbf{h}_{b,k}=\mathbf{F}\mathbf{h}_{k}\in {\mathbb{C}}^{M\times 1}$ is the beamspace channel vector between BS and the $k$-th user and $\mathbf{h}_{k}$ is the downlink channel vector in (1). The beamspace channel matrix, $\mathbf{H}_{b}$, has a sparse nature \cite{8, 9, 12}, where the number of dominant elements of each beamspace channel vector, $\mathbf{h}_{b,k}$, is much smaller than $M$. This sparse nature can be utilized to construct a low-dimensional codebook using by a single beam selection criterion ~\cite{11}. In particular, with the sparse beamspace channel matrix, only a few beams can be selected to serve $K$ users simultaneously. Therefore, we assume the $N_{RF_{g}}\times K$  dimensional multi-user equivalent beamspace channel matrix, $\mathbf{H}_{eq}$, which is given by the following equation:
\begin{equation}
 \begin{split}
 \mathbf{H}_{eq}&=[\mathbf{S}_{h}^{H}\mathbf{h}_{b,1},\mathbf{S}_{h}^{H}\mathbf{h}_{b,2},...,\mathbf{S}_{h}^{H}\mathbf{h}_{b,K}]\\
 &=[\mathbf{h}_{eq,1},\mathbf{h}_{eq,2},...,\mathbf{h}_{eq,K}],
 \end{split}
 \end{equation}
 where $\mathbf{h}_{eq,k}=\mathbf{S}_{h}^{H}\mathbf{h}_{b,k}$ is the $N_{RF_{g}}\times 1$ represents 
equivalent beamspace channel vector. Thus, multi-stream digital precoding can be applied to $\mathbf{H}_{eq}$, where simple ${N_{RF_{g}}}\times K$ dimensional hybrid ZF precoding is performed by the designed equivalent beamspace channel, $\mathbf{H}_{eq}$ as follows:
\begin{equation}
 \mathbf{W}=\mathbf{H}_{eq}^{H}\left(\mathbf{H}_{eq}\mathbf{H}_{eq}^{H}\right)^{-1}\mathbf{\Lambda},
 \end{equation}
where $\mathbf{\Lambda}$ is a diagonal matrix, introduced to transmit 
 power normalization at the $k$-th user, $\mathbf{w}_{i}=\frac{\mathbf{\hat{H}}_{eq}^{\dag}\left(:,i\right)}{\|\mathbf{\hat{H}}_{eq}^{\dag}\left(:,i\right)\|}$ is the ZF precoding vector, $\mathbf{\hat{H}}_{eq}=\left[\mathbf{\hat{h}}_{eq,k}\right]\in \mathbb{C}^{N_{RF_{g}}\times K}$, 
 and the operator $(\cdot)^{\dagger}$ is the Moore-Penrose pseudoinverse.

\section{Rate-Loss Mitigation}
In this section, we analyze the expected quantization error (QE) using the proposed HBS scheme in an equaivalent beamspace MIMO channel and qualify the rate degradation.\\
{\textit{A. The Rate Loss}:} The rate loss, $\triangle R(\rho)$ is expressed with the designed HBS scheme as
\begin{equation}
 \triangle R(\rho)= R_{pzf}(\rho)- R_{h}(\rho),
 \end{equation}
where  the per-user rate, $R_{pzf}(\rho)$ is achieved by an ideal zero-forcing precoding based on an equivalent perfect channel as  

\begin{equation}
R_{pzf}(\rho)=\mathbb{E}\left[\textrm{log}_{2}\left(1+\frac{\rho}{K}|\mathbf{h}_{eq,k}^{H}\mathbf{w}_{pzf,k}|^{2}\right)\right],
 \end{equation}
 and the per-user rate, $R_{h}(\rho)$ is achieved by the designed zero-forcing precoding (5) with a feedback-based equivalent channel as
 \begin{equation}
 R_{h}(\rho)=\mathbb{E}\left[\textrm{log}_{2}\left(1+\frac{\frac{\rho}{K}|\mathbf{h}_{eq,k}^{H}\mathbf{w}_{k}|^{2}}{1+\sum_{k\neq i}\frac{\rho}{K}|\mathbf{h}_{eq,k}^{H}\mathbf{w}_{i}|^{2}}\right)\right].
 \end{equation}
Next, by substituting (7) and (8) into (6) and using [2, Theorem 1] and [11], the resultant upper bound of the rate loss, $\triangle R(\rho)$ is given by 
\begin{equation}
\triangle R(\rho)\leq\textrm{log}_{2}\left\{1+\gamma(K-1)\mathbb{E}\left[Z_h\right]\right\},
\end{equation}
where $\mathbb{E}\left[Z_h\right]$ denotes the expected quantization error (QE) and $Z_h=1-\textrm{cos}^{2}\left(\measuredangle(\mathbf{\tilde{h}}_{eq,k},\mathbf{\hat{h}}_{eq,k})\right)$ represents the estimated QE, $\mathbf{\tilde{h}_{eq,k}}=\frac{\mathbf{h}_{eq,k}}{\|\mathbf{h}_{eq,k}\|}$ is the equivalent channel direction,  $\mathbf{\hat{h}}_{eq,k}=\|\mathbf{h}_{eq,k}\|\mathbf{\xi}_{k,\mathcal{C}_{k}}$ is an equivalent feedback channel vector, $\mathbf{\xi}_{k,i}=\mathbf{S}\mathbf{}_{h}^{H}\mathbf{Fc}_{k,i}, \forall_{k\neq i}$ is the $N_{RF_{g}}\times 1$ hybrid codeword, $\mathbf{c}_{k,i}$ is the $M\times 1$ large-dimensional channel vector, $\mathbf{\mathcal{C}}_{k}=\textrm{arg min}_{i} Z_h$ becomes feedback by $N_h\geq \frac{\gamma}{3}\left(L_h-1\right)+\left(L_h-1\right)\textrm{log}_{2}(K-1)$ bits, $\gamma=10\textrm{log}_{10}\frac{\rho}{K}\mathbb{E}\left[\|\mathbf{h}_{eq,k}\|^{2}\right]$ is the resultant received signal-to-noise ratio (SNR), $L_h$ is the number of the estimated hybrid channel propagation clusters which significantly impact on the QE, and $i\in 1,2,...,2^{N_h}$.\par

{\textit{B. The Expected Quantization Error Analysis}:}
Let $L_h=\xi L_1+(1-\xi) L_2$ where $L_1=\frac{N_{RF_{g_1}}}{K}$ and $L_2=\frac{N_{RF_{g_2}}}{K}$. Then, using [13, Lemma 1], the complementary cumulative distribution function is given by
\begin{equation}
P_{r}\left(Z_h\geq z_h\right)=\left(1-z_h^{L_h-1}\right)^{2^{N_h}}.
\end{equation}
Thus, the expected QE, $\mathbb{E}\left[Z_h\right]$, is given by (10)
\begin{equation}
\begin{split}
\mathbb{E}\left[Z_h\right]
&=\int_{0}^{1}\left(1-z_h^{L_h-1}\right)^{2^{N_h}}dz=I_{h},
\end{split}
\end{equation}
where the integral, $I_{h}$ denotes the gamma function and is given using by \cite{14}:
\begin{equation}
\begin{split}
I_{h}
&=\frac{1}{L_h-1}\beta \left(2^{N_h}+1,\frac{1}{L_h-1}\right)\\
&=\frac{2^{N_h}\Gamma\left(2^{N_h}\right)\Gamma\left(\frac{L_h}{L_h-1}\right)}{\Gamma\left(2^{N_h}+\frac{L_h}{L_h-1}\right)},
\end{split}
\end{equation}
where $\beta\left(x,y\right)=\frac{\Gamma(x)\Gamma(y)}{\Gamma(x+y)}$, $\beta\left(\cdot\right)$ denotes the beta function in [15, p.5] and $\Gamma(z_h+1)=z_h\Gamma(z_h)$ depicts the fundamental equality.\\
\textit{\textbf{Case I}:} If $\xi=0$, then the integral $I_{h}$ (12) become 
\begin{equation}
I_{h}=\frac{2^{N_2}\Gamma(2^{N_2})\Gamma\left(\frac{L_2}{L_2-1}\right)}{\Gamma\left(2^{N_2}+\frac{L_2}{L_2-1}\right)}=I_{g_2},
\end{equation}
where $N_h\geq \frac{\gamma}{3}\left(L_2-1\right)+\left(L_2-1\right)\textrm{log}_{2}(K-1)=N_2$ is the number of feedback bits. Now, if we set the size of RF chains group $2$, $N_{RF_{g_2}}=32$ and the number of the total active users, $K=16$ in (13). Thus, we have,
\begin{equation}
I_{g_2}=\frac{2^{N_2}\Gamma(2^{N_2})\Gamma\left(2\right)}{\Gamma\left(2^{N_2}+2\right)}<2^{-N_2}.
\end{equation}
From (14), we observe that $I_{g_2}$ satisfies an ideal Voronoi region around a quantization error, which is a spherical cap of area $2^{-N_2}$ in [13]. In this scenario, the beta function does not affect on path gains. Now, using (14) in (9), we obtain the rate loss for $\textbf{Case I}$ as follows:
\begin{equation}
\triangle R(\rho)\leq\textrm{log}_{2}\left\{1+\gamma(K-1)2^{-N_2}\right\}.
\end{equation}
\textit{\textbf{Case II}:} If $\xi=1$, then the integral $I_{h}$ (12) become 
\begin{equation}
I_{h} =\frac{2^{N_1}\Gamma(2^{N_1})\Gamma\left(\frac{L_1}{L_1-1}\right)}{\Gamma\left(2^{N_1}+\frac{L_1}{L_1-1}\right)}=I_{g_1},
\end{equation}
 where $N_h\geq \frac{\gamma}{3}\left(L_1-1\right)+\left(L_1-1\right)\textrm{log}_{2}(K-1)=N_1$ is the number of feedback bits. Now, if we set the number RF chains group $1$, $N_{RF_{g_1}}=48$ and the total active users $K=16$ in (16), then by applying Kershaw's law of inequality for a gamma function [16], we get from (16)
 \begin{equation}
I_{g_1} =\frac{2^{N_1}\Gamma(2^{N_1})\Gamma\left(\frac{3}{2}\right)}{\Gamma\left(2^{N_1}+\frac{3}{2}\right)}<2^{\frac{-N_1}{2}},
\end{equation}
where the gamma function $\Gamma(1)=\Gamma(2)=1$ and $\Gamma(x)\leq 1$ for $1\leq x \leq 2$. Thus, using (17) in (9), we can obtain the rate loss for $\textbf{Case II}$ as follows:
\begin{equation}
\triangle R(\rho)\leq\textrm{log}_{2}\left\{1+\gamma(K-1)2^{\frac{-N_1}{2}}\right\}.
\end{equation}
From (14) and (17), we observe that the outcome of the beta function depends on both the size of the selected beam or RF chain group and the number of total active users. Therefore, the rate-loss performance entirely rely on the size of RF chain group and the number of total active users.
Now, using (15) and (18), the rate loss per user, $\triangle R(\rho)$ can be computed as in Table I, considering $N_h$ feedback bits where $N_h$ scales linearly with $L_h-1$ to maintain a finite rate loss.

\section{Simulation Results}
In this section, we illustrate some simulation results to demonstrate the proposed HBS scheme in an equivalent beamspace MIMO downlink channel. Throughout the simulations, we assumed that $M=256$, $K=16$, $SNR=12\textrm{dB}$, $f_{c}=60$ $\textrm{GHz}$, $\lambda=5\textrm{mm}$, $d=\lambda/2$ and $10$ iterations. In a fair comparison, our HBS scheme achieved an extremely smaller rate loss per-user in both channel cluster $L=2$ and $L=3$ cases and outperformed the traditional SBS scheme (as shown in Table I). Cluster $L=1$ case exhibit a higher rate-loss ($0.73$ bps/Hz/user) due to the invalid feedback bits. Table I demonstrates the rate loss of the HBS scheme is $0.12$ bps/Hz/user, $0.10$ bps/Hz/user and $0.09$bps/Hz/user in both  propagation clusters $2$ and $3$ cases, while the rate loss of the SBS scheme is $0.34$ bps/Hz/user in the limited $L=3$ cases. This confirms the significantly lower rate loss per-user of our HBS scheme compared with that of the SBS scheme in \cite{11}.\par
Figs. 2 and 3 show per-user data rate comparisons for the proposed HBS scheme aganist the SBS setup and the random vector quantization (RVQ) scheme at the 12 dB SNR, $\gamma$. 
A zero-forcing precoding is applied in computer simulations. In Fig. 2, if $\xi=0$, $N_{RF_{g_1}}=32$ and $N_{RF_{g_2}}=16$, then $\mathbf{S}_{h}$ enable $\textbf{S}_2$ and select $N_{RF_{g_2}}$ to operate the channel cluster $L=1$ with a zero-feedback bit by each active user. Thus, the proposed HBS scheme achieves an worse rate performance because of the reduction of the diversity gain. If we consider $\xi=1$ with the aforesaid above environment of Fig. 2, then $\mathbf{S}_{h}$ enable $\textbf{S}_1$ and select $N_{RF_{g_1}}$ to operate the channel cluster $L=2$ with $7$ feedback bits by each active user. Consequently, the proposed HBS scheme achieves a huge rate performance per-active-user and outperforms the SBS scheme as shown in Fig. 2. Meanwhile in Fig.3, we increase the size of RF chain group, for example, if we consider $N_{RF_{g_1}}=48$ and $N_{RF_{g_2}}=32$ and apply similar fashion as in Fig. 2, then each active user can significantly operate the channel propagation clusters 2 and 3 considering the proposed HBS scheme and achieve an enormous rate performance of the systems. For an ideal case, $\mathbf{\tilde{h}}_{eq,k}=\mathbf{\hat{h}}_{eq,k}$ to represent an error-free feedback channel.
  \begin{table*}
	\centering
	\caption{\small {\large{C}\small {OMPUTATION} OF THE \large{R}\small{ATE}-LOSS \large{P}\small{ER-USER}}}
	\begin{tabular}{|*{10}{p{11.45cm}|}}
		\hline
		\centering
		\small The rate loss/user, $\triangle R(\rho)$\\
		(\small $M=256, K=16$, $\textrm{SNR}, \gamma=12 \textrm{dB}$, $iterations=10$)
	\end{tabular}\\	
	\begin{tabular}{|c|c|c|c|}
\hline
\multirow{1}*[-.1ex]{\small Scheme}&\multicolumn{1}{|c|}{\small Rate-loss [bps/Hz/user]}\\\cline{1-2}
\small  SBS Scheme [11] &0.34, if $N_{RF}=48$, $L=3$, $N=15$\\ \hline
\small &0.73, if $N_{RF_{g_1}}=32$, $N_{RF_{g_2}}=16$, $\xi=0$, $L_2=1$, $N_2=0$\\ \cline{2-1}
\small Proposed&0.12, if $N_{RF_{g_1}}=32$, $N_{RF_{g_2}}=16$, $\xi=1$, $L_1=2$, $N_1=7$\\\cline{2-1}
\small HBS Scheme&0.10, if $N_{RF_{g_1}}=48$, $N_{RF_{g_2}}=32$, $\xi=0$, $L_2=2$, $N_2=7$\\ \cline{2-1}
\small &0.09, if  $N_{RF_{g_1}}=48$, $N_{RF_{g_2}}=32$, $\xi=1$, $L_1=3$, $N_1=15$\\ \hline
\end{tabular}\\
\end{table*}
\raggedbottom
 However, a huge rate loss in~\cite{2} and a considerable rate loss in~\cite{11} are observed. Lastly, the per-user data rate of each scheme rises with an increasing SNR, as expected.
 \begin{figure}[t]	
  \centering
  \includegraphics[width=3in,height=2.1in,keepaspectratio]{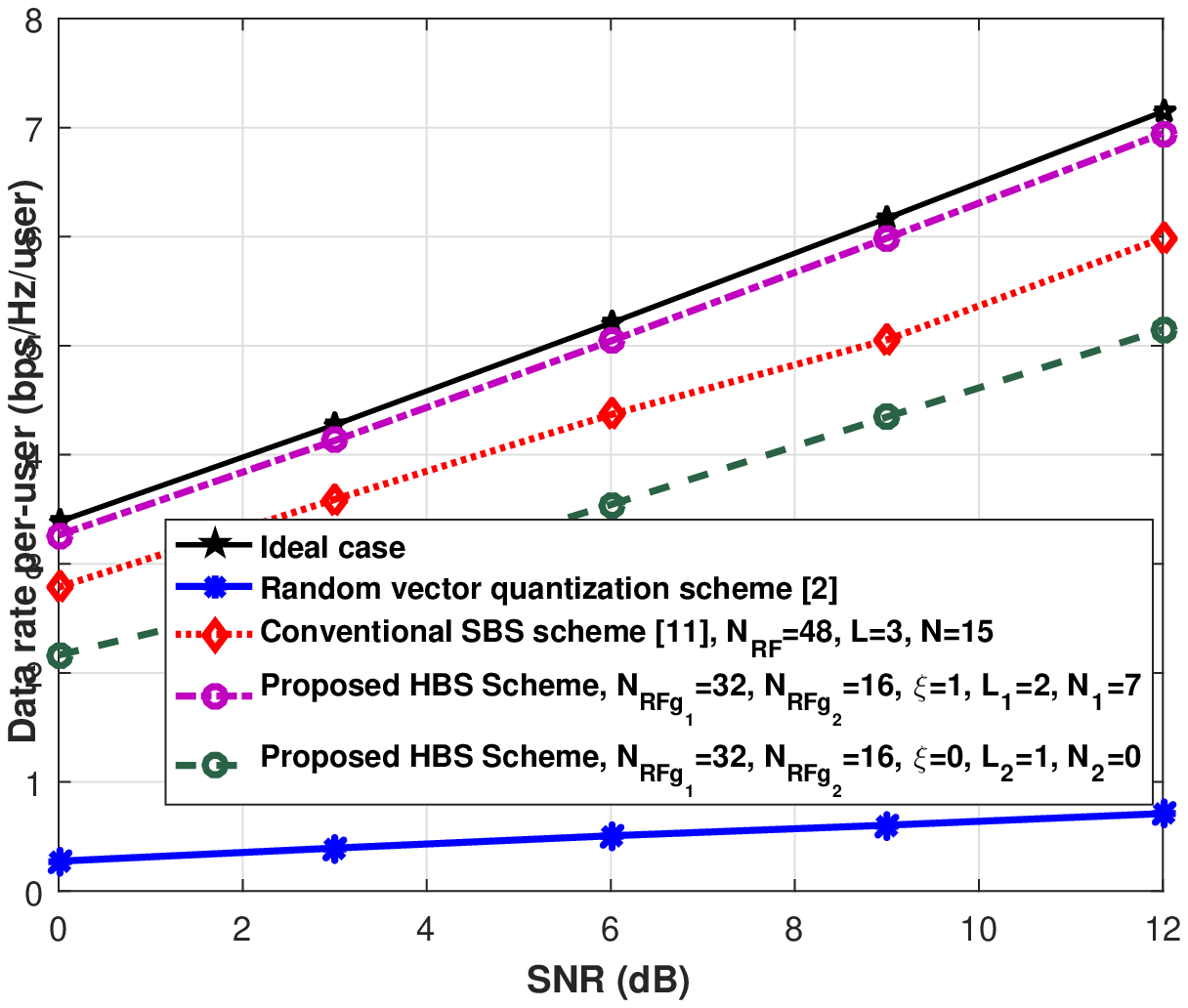}
  \caption{\small Per-user data rate comparison with $K=16$ users. $L=3$, $N=15$ for [11], $\xi=0$, $L_2=1$, $N_2=0$ for the proposed HBS \textbf{Case I} and $\xi=1$, $L_1=2$, $N_1=7$ for the proposed HBS \textbf{Case II}.}
  \label{Fig2}
  \end{figure}
  \begin{figure}[t]	
   \centering
  \includegraphics[width=3in,height=2.1in,keepaspectratio]{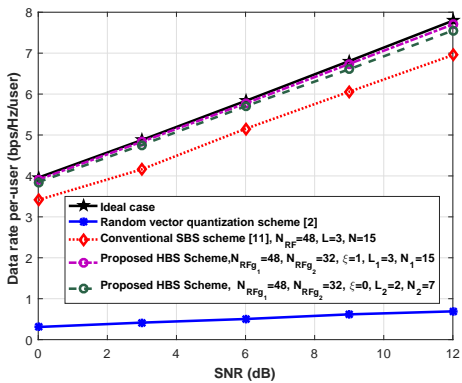}
  \caption{\small Per-user data rate comparison with $K=16$ users. $L=3$, $N=15$ for [11], $\xi=0$, $L_2=2$, $N_2=7$ for the proposed HBS \textbf{Case I} and $\xi=1$, $L_1=3$, $N_1=15$ for the proposed HBS \textbf{Case II}.}
   \label{Fig3}
   \end{figure}
   \raggedbottom

   \section{Conclusions}
 We proposed an HBS scheme to make a multiple beam or RF chain group selection opportunity and sort out a suitable channel cluster for every supported user in  mmWave beamspace MIMO LAA systems. Our HBS scheme can easily find out a reliable RF chain group and determine the minimum QE significantly. 
 The SBS scheme provides a less desirable rate performance per-active-user than does our HBS scheme. Simulation results also confirm that the proposed HBS scheme outperformed the SBS scheme in terms of data rate. Lastly, the proposed HBS scheme can be extended further to next-generation MIMO non-orthogonal multiple-access (MIMO-NOMA) networks, which will be subjects of future studies.

  \raggedbottom

 \ifCLASSOPTIONcaptionsoff
    \newpage
  \fi

  \end{document}